\documentclass[usenatbib]{mn2e}
\usepackage{graphicx,natbib,amssymb,amsmath}
\usepackage{txfonts, stfloats}
\usepackage{siunitx}
\usepackage{xcolor}
\usepackage{color}
\usepackage{soul}

\usepackage{hyperref}

\usepackage{xspace}
\newcommand{\FIRAS}{{\it COBE/FIRAS}\xspace}

\newcommand{\PIXIE}{{\it PIXIE}\xspace}
\newcommand{\SuperPIXIE}{{\it SuperPIXIE}\xspace}
\newcommand{\Voyage}{{\it Voyage~2050}\xspace}

\newcommand{\Planck}{{\it Planck}\xspace}

\newcommand{\omrel}{\Omega_{\rm rel}}

\newcommand{\SDGWtool}{\href{https://github.com/CMBSPEC/GW2SD.git}{\tt GW2SD}}

\newcommand{\pot}[2]{#1 \times 10^{#2}}
\newcommand{\zmax}{{z_{\rm max}}}
\newcommand{\Mpc}{{{\rm Mpc}}}
\newcommand{\Hz}{{{\rm Hz}}}

\graphicspath{{./Immagini/}}

\newcommand{\pd}[2]{\frac{\partial{#1}}{\partial{#2}}}
\newcommand{\diff}{\mathop{}\!\mathrm{d}}

\definecolor{light-gray}{gray}{0.70}

\title[Bridging the gap]{Bridging the gap: spectral distortions meet gravitational waves}

\date{\vspace{-3mm}{Accepted 2020 --. Received 2020 --}}

\begin{document}

\author[Kite et al.]{
Thomas Kite$^{1}$\thanks{E-mail: \href{mailto:thomas.kite@manchester.ac.uk}{thomas.kite@manchester.ac.uk}},
Andrea Ravenni$^{1}$\thanks{E-mail: \href{mailto:andrea.ravenni@manchester.ac.uk}{andrea.ravenni@manchester.ac.uk}},
Subodh P. Patil$^{2}$\thanks{E-mail: \href{mailto:patil@lorentz.leidenuniv.nl}{patil@lorentz.leidenuniv.nl}}
and Jens Chluba$^{1}$\thanks{E-mail: \href{mailto:jens.chluba@manchester.ac.uk}{jens.chluba@manchester.ac.uk}}
\\
$^1$Jodrell Bank Centre for Astrophysics, School of Physics and Astronomy,
The University of Manchester, Manchester, M13 9PL, U.K.
\\
$^2$Instituut-Lorentz for Theoretical Physics, Leiden University, 2333 CA Leiden, The Netherlands
}

\maketitle

\begin{abstract}
Gravitational waves (GWs) have the potential to probe the entirety of cosmological history due to their nearly perfect decoupling from the thermal bath and any intervening matter after emission. In recent years, GW cosmology has evolved from merely being an exciting prospect to an actively pursued avenue for discovery, and the early results are very promising. As we highlight in this paper, spectral distortions (SDs) of the cosmic microwave background (CMB) uniquely probe GWs over six decades in frequency, bridging the gap between astrophysical high- and cosmological low-frequency measurements. This means SDs will not only complement other GW observations, but will be the sole probe of physical processes at certain scales. To illustrate this point, we explore the constraining power of various proposed SD missions on a number of phenomenological scenarios: early-universe phase transitions (PTs), GW production via the dynamics of SU(2) and ultra-light U(1) axions, and cosmic string (CS) network collapse. We highlight how some regions of parameter space were already excluded with data from \FIRAS, taken over two decades ago. To facilitate the implementation of SD constraints in arbitrary models we provide \SDGWtool. This tool calculates the window function, which easily maps a GW spectrum to a SD amplitude, thus opening another portal for GW cosmology with SDs, with wide reaching implications for particle physics phenomenology.
\end{abstract}


\section{Introduction}
Gravitational wave (GW) astronomy has become a reality. 
The now routine detection of compact object mergers by the LIGO/Virgo collaboration \citep{LIGOScientific:2018mvr} has made, for good reasons, the study of GWs one of the most active and current topics in cosmology and astrophysics.
Ongoing and planned observations of the tensor perturbation power spectrum currently span some 21 orders of magnitudes of frequency: From cosmic microwave background (CMB) upper limits on primordial B-modes \citep{Ade:2018gkx, Aghanim:2018eyx} measurements at the lowest frequencies, to interferometry detections of GWs \citep[e.g.,][]{Abbott2020ApJ, Abbott2020II} and Pulsar Timing Array (PTA) measurements \citep[e.g.,][]{Perera:2019sca, Alam2020NG} at higher frequencies. In the next few years, a plethora of experiments will test different scales between these extremes \citep[e.g.,][for overview]{Campeti:2020xwn}.

Many physical processes can indeed lead to detectable tensor perturbations \citep[see][for review]{Caprini:2018mtu}. These include GWs from phase transitions \citep{Caprini:2018mtu, Nakai2020}, early universe gauge field production \citep{ Dimastrogiovanni:2016fuu, Machado:2018nqk, Machado:2019xuc}, and cosmic string networks \citep{Buchmuller:2019:CSGW}.
Given these exciting theoretical developments, it is interesting to ask which cosmological and astrophysical probes can help constrain these different scenarios. In this paper, we show that CMB spectral distortions (SDs) can provide complementary information at frequencies \mbox{$f=10^{-15}$--$10^{-9}\,\Hz$} unavailable to other probes. In this way, SDs offer a bridge between scales probed by next-generation CMB surveys \citep[e.g.,][]{Ade:2018sbj, Hazumi:2019lys, Delabrouille:2019thj}, and astrophysical GW observatories such as current \citep[e.g.][]{Perera:2019sca} and future \citep[e.g.][]{Bull:2018lat} PTA measurements.

How do CMB SDs constrain tensor perturbations at the scales that they do? Spectral distortions are created by mechanisms that lead to energy release into the photon-baryon fluid at redshifts $z\lesssim 2\times 10^6$, when thermalization processes cease to be efficient \citep{Zeldovich:1969ff, Sunyaev:1970er, Illarionov1975, Danese1982, Burigana1991, Hu1993, Chluba2011therm}.
Many sources of distortions exist within standard $\Lambda$CDM cosmology as well as scenarios invoking new physics \citep[see][for broad overview]{Chluba2019BAAS}, and innovative experimental concepts \citep{Kogut2016SPIE, Kogut2019BAAS, Chluba2019Voyage} have now reached critical thresholds to significantly advance the long-standing distortion constraints from \FIRAS \citep{Mather1994, Fixsen1996}.
A particular source of SDs is due to the dissipation of tensor modes while they travel almost unimpeded through the cosmic plasma \citep{Ota2014, Chluba2015}.

How do tensor perturbations distort the CMB spectrum? In general, perturbations in the photon fluid dissipate through electron scattering and free-streaming effects. Dissipation of scalar perturbations provides one of the guaranteed sources of SDs in the early Universe within the standard thermal history \citep[e.g.,][]{Sunyaev1970diss, Daly1991, Hu1994, Chluba2012, Chluba2012inflaton}.
Similarly, tensor modes lose a small fraction of their energy by continuously sourcing perturbations in the photon fluid which then also distort the CMB spectrum. In contrast to scalar modes, however, the dissipation is mainly mediated by free-streaming effects. As shown in detail by \citet{Chluba2015}, this leads to dissipation of perturbations over a vast range of scales, extending far beyond those relevant to scalar perturbations. Thus, although the tensor dissipation rate is suppressed relative to scalar dissipation (tensor modes are not significantly damped by interactions with the photons), this opens new avenues for model constraints from SDs.

Building on \cite{Chluba2015}, we translate the relations between $\mu$-distortions and primordial tensor perturbation into quantities commonly used for GW searches. This makes it easier to compare SD limits to those from other probes.
As examples we consider several inflationary models which source GWs beyond vacuum fluctuations, early-universe phase transitions (PTs) and cosmic string (CS) networks, all of which demonstrate how SD measurements are and will be important for excluding portions of their respective parameter spaces.
Indeed we highlight that several of the widely discussed models could have already constrained some regions of their respective parameter spaces with SD limits from \FIRAS, taken over a quarter-century ago. Future spectrometer concepts like \PIXIE \citep{Kogut2016SPIE} and its enhanced versions \citep[e.g.,][]{PRISM2013WPII, Kogut2019BAAS} could, through their increased sensitivity, significantly increase the range of scales and parameter space covered. This could give CMB spectral distortions an important role in this highly-synergistic multi-messenger campaign, providing unique scientific opportunities for the next generation of cosmologists and particle phenomenologists alike.

\vspace{-5mm}
\section{GWs in the expanding Universe}
A GW can be represented as a transverse traceless tensor perturbation of the metric's spatial component, $h_{ij}$, and the energy density it carries is $\rho_\text{GW} = \langle h'_{ij} h'^{ij} \rangle / (32 \pi G)$, where the prime denotes conformal time derivatives\footnote{We adopt the normalization conventions of \citet{Watanabe:2006qe}.}. If these GWs were produced primordially\footnote{We consider the case of sub-horizon generation further on.}, we can define the GW fractional energy density per decade of wavelengths as \citep[e.g.,][]{Watanabe:2006qe}
\begin{equation}
    \Omega_\text{GW}(k,\eta)
    = 
    \frac{1}{\rho_{\rm c}(\eta)}
    \pd{\rho_\text{GW}(k,\eta)}{\ln k}
    =
    \frac{\mathcal{P}_T(k)}{12 a^2(\eta) H^2(\eta)}
    \left[
        \mathcal{T}_\text{GW}'(k,\eta)
    \right]^2 ,
\label{eq:OmegaGW_PT_relation}
\end{equation}
where $\rho_{\rm c}$ is the critical density, and in the second equality we factored the primordial tensor power spectrum $\mathcal{P}_T$ and the deterministic GW transfer function $\mathcal{T}_\text{GW}$.

In \cite{Watanabe:2006qe}, several analytical approximations of the GW transfer function were developed.
During radiation domination (RD) we have
\begin{equation}
    \left[\mathcal{T}'_\text{GW}(k,\eta)\right]^2
    \approx
    k^2
    \left[
        j_1(k \eta)
    \right]^2,
\label{eq:GW_Transf_RAD}
\end{equation}
whereas during matter domination (MD), one finds
\begin{gather}
    \left[\mathcal{T}'_\text{GW}(k,\eta)\right]^2
\approx
    \begin{cases}
       k^2 \frac{\eta_\text{eq}^2}{\eta^2}
       \left[
            A(k) j_2(k \eta)
            + B(k) y_2(k \eta)
       \right]^2 & \text{if } k > k_\text{eq}
     \\
       k^2
       \left[
            \frac{3 j_2(k \eta)}{k\eta}
       \right]^2 & \text{if } k < k_\text{eq}
     \end{cases}\,,
\label{eq:GW_Transf_MAT}
\nonumber\\
    A(k)
    =
    \frac{3}{2 k \eta_\text{eq}}
    - \frac{\cos (2 k \eta_\text{eq})}{2 k \eta_\text{eq}}
    + \frac{\sin (2 k \eta_\text{eq})}{(k \eta_\text{eq})^2} \, ,
\\ \nonumber
    B(k)
    =
    -1 + \frac{1}{(k \eta_\text{eq})^2}
    - \frac{\cos (2 k \eta_\text{eq})}{(k \eta_\text{eq})^2}
    - \frac{\sin (2 k \eta_\text{eq})}{2 k \eta_\text{eq}} \, .
\end{gather}
Here, $k_\text{eq}$ is the comoving wavenumber entering the horizon at the time of matter-radiation equality $\eta_\text{eq}$%
, and $j_\ell$ and $y_\ell$ are the spherical Bessel functions of first and second kind.
For wavelengths much smaller than those entering the horizon today ($k\eta_0\gg1$) we can expand the GW transfer function derivatives at leading order in $k$.
Additionally, since we always observe quantities that involve $(\mathcal{T}_\text{GW}')^2$ integrated over some range of $k$, we can average over one period to obtain \citep[e.g.,][]{Caprini:2018mtu}
\begin{equation}
    \langle \left[\mathcal{T}'_\text{GW}(k,\eta) \right]^2\rangle 
    \stackrel{\stackrel{k\eta_0\gg 1}{\downarrow}}{\approx}
    \eta_\text{eq}^2/2\eta^4\, ,
\label{eq:MD_single_cycle}
\end{equation}
which is a smooth function of $k$ valid during MD.
Similarly, during RD we can apply the same procedure to Eq. \eqref{eq:GW_Transf_RAD}, and obtain
\begin{equation}
    \langle \left[\mathcal{T}'_\text{GW}(k,\eta)\right]^2 \rangle
    \approx
    1/2\eta^2 \, .
\label{eq:RD_single_cycle}
\end{equation}
For later use we point out that during RD, where Eq. \eqref{eq:RD_single_cycle} is valid, $a \propto \eta$, while during MD relevant to Eq.~\eqref{eq:MD_single_cycle} we have $a\propto \eta^2$. Together with Eq.~\eqref{eq:OmegaGW_PT_relation}, this means that the GW energy density at a given scale evolves as $\Omega_{\rm GW}\propto a^{-4} H^{-2}\propto {\rm const}$ during RD and $\Omega_{\rm GW}\propto a^{-4} H^{-2}\propto (1+z)$ in the MD era.

As pointed out in \cite{Watanabe:2006qe}, the approximations given above neglect some important details. One of these is the process of neutrino damping, which has its greatest effects on scales important to SD physics. The damping is effective during RD but only after neutrino decoupling $(T\lesssim 2\text{MeV})$, which taken together almost exactly coincides with the SD regime. This damping occurs since free streaming neutrinos correspond to a non-negligible fraction of the energy density of the Universe during RD and generate significant anisotropic stresses that result in the damping of tensor perturbations. The magnitude of the effect is a $35.6\%$ decrease of the power available in GW \citep{Weinberg:2004:nuDamp}. To include this effect the transfer function given in \cite{Dicus:2004:nuDamp} is used:
\begin{equation}
    \mathcal{T}_{\rm GW}'=
    \frac{1}{\eta}\sum_{n\,\,\text{even}}
    a_n\left[
        n j_n(k\eta) - k\eta j_{n+1}(k\eta)
    \right],
\end{equation}
with the coefficients $a_0=1$, $a_2=0.243807$, $a_4=5.28424\times 10^{-2}$ and $a_6=6.13545\times 10^{-3}$. This is valid for the range of scales needed in the following section.

It is clear from Eqs. (\ref{eq:OmegaGW_PT_relation}), (\ref{eq:MD_single_cycle}) and (\ref{eq:RD_single_cycle}) that the exact transfer functions are important quantities for comparing the effects of the GW background in the early and late Universe. In this paper, we compare the SD sensitivity ($\leftrightarrow$ early Universe) to PTA and interferometry ($\leftrightarrow$ late Universe). Even CMB temperature anisotropies, although sourced early on, mostly probe the Universe after the RD-MD transition. Because of this it is essential to get the exact dynamics of this transition right for any comparison to be meaningful. To study the evolution of the GW background in detail we numerically solved for the wave evolution through the RD-MD transition, which gave results agreeing with \cite{Watanabe:2006qe} and \cite{Dicus:2004:nuDamp} in the appropriate limits, while allowing us to more carefully model the GW background for a realistic cosmology involving neutrino and dark energy densities.

For the \Planck 2018 best-fit cosmology \citep{Aghanim:2018eyx}, the exact solution is well approximated by\footnote{This does \textbf{not} include a factor of $1/2$ for oscillations}
\begin{equation}
    \Omega_{\rm GW}/\mathcal{P}_T = \frac{\mathcal{D} \omrel}{12}\left(1 + \alpha_1\kappa^{-1} + \alpha_2\kappa^{-3/2} + \alpha_3\kappa^{-2}\right),
    \quad \kappa = \frac{k}{k_*},
\end{equation}
where $\alpha_1 = 5.74$, $\alpha_2 = -2.47$, $\alpha_3 = 14.48$, $k_*=1/551 \text{ Mpc}^{-1}$, $\mathcal{D} = 0.642$ is the neutrino damping factor and $\omrel = 9.19 \times 10^{-5}$ is the combined density of radiation and neutrinos, treating the latter as massless.
This solution differs by $\sim 20\%$ from \cite{Caprini:2018mtu}, with better matching arising when neglecting the neutrino energy density. For details see Kite et. al. (in preparation).

Since gravitational wave upper limits are usually quoted as function of frequencies rather than wavelengths, we will use the relation
$
    k / \text{Mpc}^{-1}
    =
    6.5 \times 10^{14}
    f / \text{Hz}
$
 to change units.

\vspace{-2mm}
\section{$\mu$-distortions from tensor perturbations}

Much like scalar perturbations, tensor perturbations dissipate over time, both transferring energy to neutrinos and (in smaller proportion) to photons. Primordial tensor perturbations entering the horizon during or slightly before the $\mu$-era ($\pot{5}{4}\lesssim z\lesssim\pot{2}{6}$), when dissipating, generate $\mu$-distortions of the CMB that will be observable today \citep{Ota2014, Chluba2015}.

The average value of $\mu$-distortions today is related to the primordial tensor power spectrum via
a window function $W_\mu (k)$
\begin{equation}
    \langle \mu_\text{GW} \rangle (\eta_0)
    =
    \int \diff \ln k \,
    W_\mu (k)\,
    \mathcal{P}_T(k) \, ,
\label{eq:mu_GW}
\end{equation}
which already averages oscillations by integrating over a transfer function's evolution throughout the $\mu$-era, achieving the usual factor of $1/2$ implicitly.
We calculate $W_\mu (k)$ numerically according to \cite{Chluba2015}. The window function is shown in Fig.~\ref{fig:z_max_window}. In comparison to the corresponding $k$-space window function of scalar perturbations \citep[e.g.,][]{Chluba2012inflaton, Chluba2015IJMPD}, the dissipation efficiency of tensors is about five orders of magnitude smaller, highlighting how weakly tensor modes couple to the photon fluid.

Offsetting this loss, we can see that tensor modes contribute to the generation of $\mu$-distortions over a vast range of scales, with a power-law decay of contributions at $k\gtrsim 10^{6}\,{\rm Mpc}^{-1}$ (Fig.~\ref{fig:z_max_window}). This is in stark contrast to the dissipation of scalar perturbations, which are limited to scales $k\simeq 50-10,000\,{\rm Mpc}^{-1}$, with a strong exponential decay of contributions from $k\gtrsim 10,000\,{\rm Mpc}^{-1}$ \citep[e.g.,][]{Chluba2012inflaton}.
Scalar modes damp by photon diffusion, which virtually erases all perturbations once the dissipation scale is crossed. For tensors, the photon damping is minute and photon perturbations are continuously sourced by the driving tensor force, explaining this significant difference \citep{Chluba2015}. 
This makes SDs a potentially unique probe of GWs from early-universe physics.

\vspace{-2mm}
\subsection{Time-dependent injection}
\label{sec:time_dep_injection}
Equation~\eqref{eq:mu_GW} determines the SD signal from primordial perturbations that were created during inflation and only later enter the horizon to dissipate their energy. Another possibility is to have perturbations created on sub-horizon scales at later times. This requires a generalisation of the window function formalism to account for the new time dependence. 

An immediate difference for sub-horizon injection is that neutrino damping will not occur, as this only matters for GWs that cross the horizon between neutrino decoupling and the start of MD. This means one can use the simpler versions of the transfer function, valid in RD, given in Eqs. (\ref{eq:GW_Transf_RAD}) and (\ref{eq:RD_single_cycle}). The time dependence --- which before was included in the physics underlying the window function --- has to be made more explicit.
Using redshift $z$ to better match \cite{Chluba2015}, Eq. \eqref{eq:mu_GW} can be generalized to
\begin{equation}
    \langle \mu_\text{GW} \rangle (z=0)
    =
    \int_0^\infty
        \diff \ln k 
    \int_{0}^{\infty} \diff z \,
    \mathcal{W}_\mu (k, z)\,
    \mathcal{P}_T(k, z) \, ,
\label{eq:mu_GW_time_dep}
\end{equation}
\begin{figure}
\centering 
\includegraphics[ width=\columnwidth]{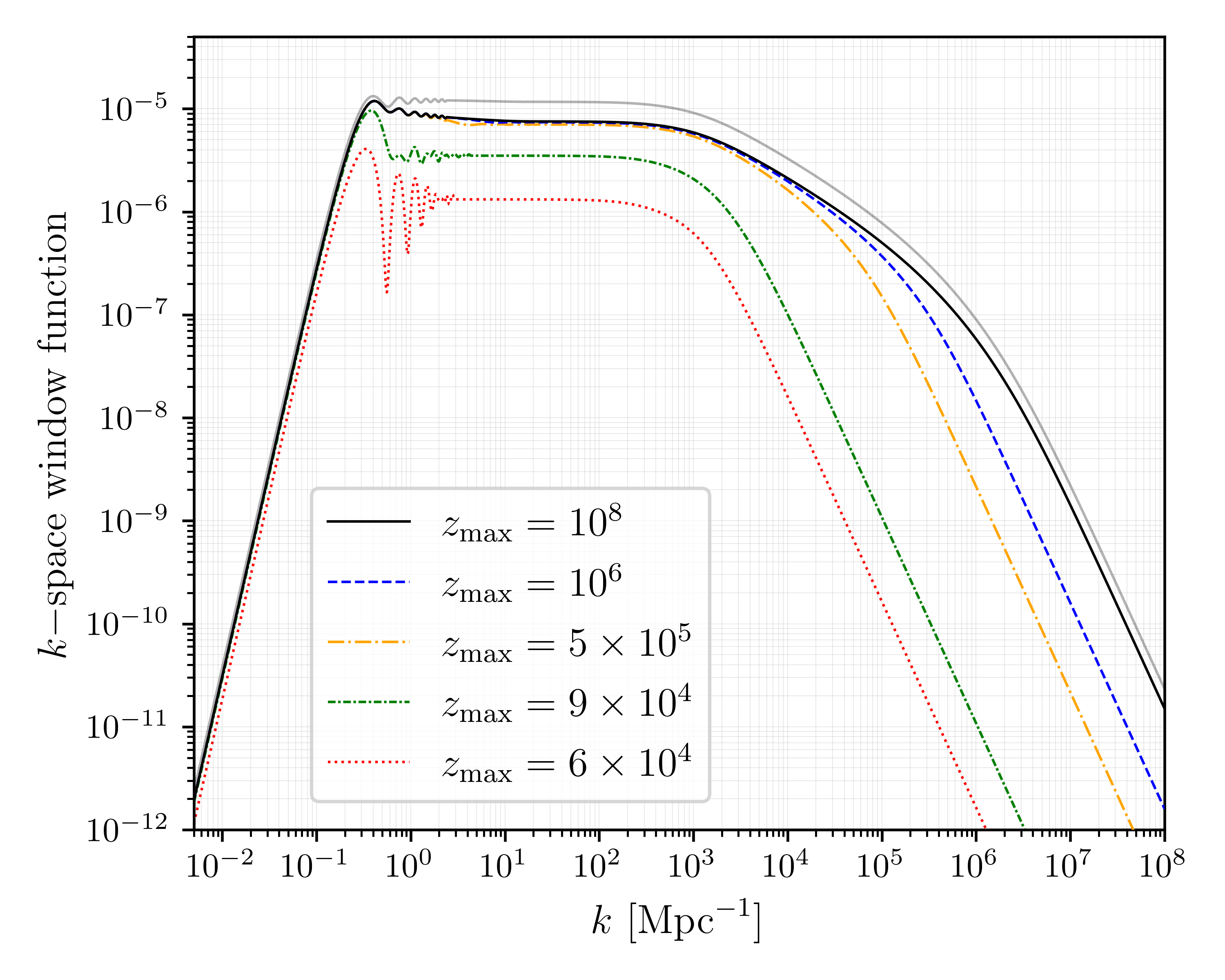}
\vspace{-7mm}
\caption{A series of curves demonstrating the form of the $k$-space window function $W^{z_{\rm max}}_\mu$ for various upper limits in redshift. For practical purposes, $z_{\rm max}=10^8$ is equivalent to $W_\mu^{\infty}\equiv W_\mu$. The solid curve is the result for power injection before the $\mu$ era, while the other curves suffer from some reduced visibility at higher $k$. The faded line shows the results without neutrino damping, leading to a $\simeq 30\%$ increase across the window function.}
\label{fig:z_max_window}
\vspace{-3mm}
\end{figure}
where we introduced the GW-$\mu$-distortion \emph{window primitive} $\mathcal{W}_\mu (k, z)$, which captures the physics behind the damping of GWs.
Note that with $\mathcal{P}_T(k,z)=\mathcal{P}_T(k)$ we recover Eq.~\eqref{eq:mu_GW} by defining $\int_0^\infty \mathcal{W}_\mu \diff z=W_\mu$. The explicit form of the window primitive is\footnote{We match the notation of \cite{Chluba2015} with $\mathcal{T}_{\rm GW}\equiv\sqrt{2}\eta\mathcal{T}_h$}.
\begin{equation}
  \mathcal{W}_\mu = 1.4\times\frac{8H^2\eta^2}{45\dot{\tau}}\left[\mathcal{T}_{\rm GW}'(k,z)\right]^2\mathcal{T}_\Theta(k,z) \,{\rm e}^{-\Gamma_\gamma^*\eta}\,\mathcal{J}_\mu(z) \, ,
    \label{eq:window_primative}
\end{equation}
and for convenience we summarize here the quantities which are relevant to calculate the window function  \citep[for their derivation and further explanation we refer to][]{Chluba2015}:
$\dot{\tau}$ is the time derivative of the Thomson optical depth. The terms $\mathcal{T}_\Theta {\rm e}^{-\Gamma_\gamma^*\eta}$ contain the physics of how the GW transfer function $\mathcal{T}_{\rm GW}$ couples to the photon fluid. These terms can be reliably approximated as
\begin{subequations}
\begin{gather}
    \mathcal{T}_\Theta(k,z)\approx\mathcal{T}_\Theta(\xi)\approx
    \frac
    {1+4.48\xi^2+91\xi^4}
    {1+4.64\xi+90.2\xi^2+100\xi^3+55\xi^4}\, ,\\
    e^{-\Gamma_\gamma^*\eta}\approx 1\, ,
\end{gather}
\end{subequations}
with $\xi=k/\tau'$. The final term $\mathcal{J}_\mu(\eta)$ is the energy branching ratio, which gives the fraction of total energy injected into the photon fluid that contributes to the $\mu$ distortion. We use the simple analytic approximation of the branching ratio \citep[`method B' in][]{Chluba:2016:SDLCDM}:
\begin{equation}
\mathcal{J}_\mu(z)\approx
\begin{cases}
{\rm e}^{-\left(z/z_{\rm th}\right)^{5/2}}
 & \text{for} \,\, z>5\times 10^{4}\\
0  & \text{otherwise}
\end{cases}
\, ,
\label{eq:branching_ratio}
\end{equation}
with $z_{\rm th}=1.98\times 10^6$ denoting the redshift where thermalisation becomes inefficient \citep[see also][]{Hu1993}.

We employ one further approximation in assuming the injection happens instantaneously across all scales at a time $\eta_*$. Therefore, the tensor perturbations are uncorrelated ($\mathcal{P}_T(k,\eta<\eta_*)=0$) up to $\eta_*$ when their power spectrum abruptly jumps to some value that we will now determine. The spectrum is found at all times after $\eta_*$ by redshifting $\Omega_\text{GW}$ from the present-day value
\begin{equation}
    \Omega_\text{GW} (k, \eta)
    =
    \begin{cases}
        \frac{a^{-4}(\eta)}{E^2(\eta)}
        \,\Omega_\text{GW} (k, \eta_0)
        &
        \eta > \eta_*
    \\
        0 
    & 
        \eta < \eta_*
    \\
    \end{cases}\,,
    \;\,
    E^2(\eta) \equiv \frac{H^2(\eta)}{H_0^2} \, ,
\end{equation}
where we used $a^{-4}/E^2 \propto \bar{\rho}_{\rm GW}/\rho_{\rm c}$.

We will only consider power injection in the RD era, hence the tensor power spectrum is then obtained using Eq.~\eqref{eq:OmegaGW_PT_relation} together with 
Eq.~\eqref{eq:GW_Transf_RAD}, and reads
\begin{equation}
    \mathcal{P}_T (k,\eta)
    =
    \begin{cases}
        \frac{12 H_0^2}{a^2k^2 [j_1(k \eta)]^2}
        \,\Omega_\text{GW}(k, \eta_0)
        &
        \eta > \eta_*
    \\
        0 
    & 
        \eta < \eta_*
    \\
    \end{cases}
\label{eq:PT_PhTr}
\end{equation}

Notice that if not for the fact that the tensor perturbations appear at $\eta_*$, the power spectrum would always be time-independent: it is in fact the equivalent of the primordial one in the ``standard'' scenario described in \cite{Chluba2015}. 

Using that Eq.~\eqref{eq:PT_PhTr} is independent of time during RD and inserting this into Eq. \eqref{eq:mu_GW_time_dep}, we can remove the time dependence in the integrand, leaving only changes in the upper limit of integration. It is therefore sufficient to study a series of window functions $W^{z_{\rm max}}_\mu=\int_0^{\zmax} \mathcal{W}_\mu \diff z$ for different upper limits in time. Examples of $W_\mu^{z_{\rm max}}$ are shown in Fig. \ref{fig:z_max_window}. We can observe that even modes originating from $z\gg 2\times 10^6$ contribute to the generation of distortions. The flat plateau of the window function at $k\simeq 0.1-10^3\,{\rm Mpc}^{-1}$ is not affected until $\zmax\lesssim \pot{5}{5}$, and will rapidly approach $W^{z_{\rm max}}_\mu\simeq0$ as $\zmax$ approaches $\pot{5}{4}$. 

For numerical applications, it is convenient to pre-tabulate the tensor window function $W^{z_{\rm max}}_\mu$  across $k$ and injection redshift $\zmax$. Since the background cosmology is fixed to high precision \citep{Planck2018params}, this procedure avoids additional approximations.\footnote{A simple interpolation routine to calculate $W^{z_{\rm max}}_\mu$ is available here: \url{https://github.com/CMBSPEC/GW2SD.git}} However, a few comments are in place: we can further improve the treatment of the transition between the $\mu$ and $y$-distortion eras, which here we modelled as a step-function [see Eq.~\eqref{eq:branching_ratio}]. Including the more gradual transition \citep[e.g., see discussion in][]{Chluba:2016:SDLCDM}, enhances the contributions from the largest scales ($k\lesssim 10^{-2}\,\Mpc^{-1}$), however, a more accurate treatment of transfer effects is also required and left to future work.

With the procedure outlined in this section we can calculate the tensor dissipation contribution to the present day value of $\mu$-distortions.
However, other processes, such as dissipation of acoustic modes and Compton cooling also source $\mu$-distortions, henceforth referred to as $\mu_\text{other}$.
Any non-detection of an enhanced level of SD would straightforwardly constrain models that generate a large $\mu_\text{GW}$, comparable to or greater than $\mu_\text{other}$.
However, things are more delicate when $\mu_\text{GW}$ becomes much smaller than the value of $\mu_\text{other}$ expected in the standard cosmological model, $\mu_{\rm other}\simeq 2\times 10^{-8}$ \citep[e.g.,][]{Chluba:2016:SDLCDM}.
In this regime, any actual analysis would require a marginalization of other sources, that we do not take into account here. However, assuming standard slow-roll inflation, we can in principle accurately predict the expected standard contribution given the power spectrum parameters measured at large angular scales \citep{Chluba2012, Khatri2013forecast, Chluba2013PCA, Cabass2016, Chluba:2016:SDLCDM}. %
For simplicity, we shall thus assume perfect removal of other $\mu$-contributions.
\newpage
Below we will consider the upper limit on $\mu$-distortions set by \FIRAS ($\mu < 9\times10^{-5}$ 95\%CL) \citep{Mather1994, Fixsen1996}, and the forecasted constraints for \PIXIE ($\mu < 3\times10^{-8}$) \citep{Kogut:2011xw}, \SuperPIXIE ($\mu < 7.7\times10^{-9}$) \citep{Kogut2019BAAS}, \Voyage ($\mu < 1.9\times10^{-9}$) and $10\times$\Voyage ($\mu < 1.9\times10^{-10}$) \citep{Chluba2019BAAS}, all of which already account for the presence of foregrounds following \citet{Abitbol2017}.

\vspace{-3mm}
\subsection{Scalar contributions}
Above we discussed separating $\mu_{\rm GW}$ from $\mu_{\rm other}$, taking the latter to be the standard model expected value. A second discussion is necessary, however, regarding the contribution that scalar perturbations have to a $\mu$ signal. This is important, considering that energetic early-universe  phenomena have the potential to generate scalar perturbations as well as tensors, which will enhance the SD production. In the following section we will discuss the scalar contributions for models where it is possible to do so, but some statements apply in general: \cite{Chluba2015} show that the corresponding window functions for scalar perturbations peak around $10^5$ higher than for tensors, but for a narrower range of scales ($k\simeq 50-10,000\,{\rm Mpc}^{-1}$ or $f\simeq \pot{8}{-14}-\pot{1.5}{-11}\,{\rm Hz}$, as previously discussed). Thus, for tensor perturbations to dominate the spectral distortion signal the scalar spectrum created must be less than $1$ part in $10^5$ of the tensor spectrum, or must be injected on smaller scales than $k\sim10^4{\rm Mpc}^{-1}$. Provided both the wider tensor window, and that some early processes will be almost invisible to scalar probes, the machinery explained above for constraining early tensor energy injection are still of interest and importance, despite the relatively low sensitivity.

\vspace{-5mm}
\section{Minimally parametric constraints}
\begin{figure*}
\centering 
\includegraphics[width=2.12\columnwidth]{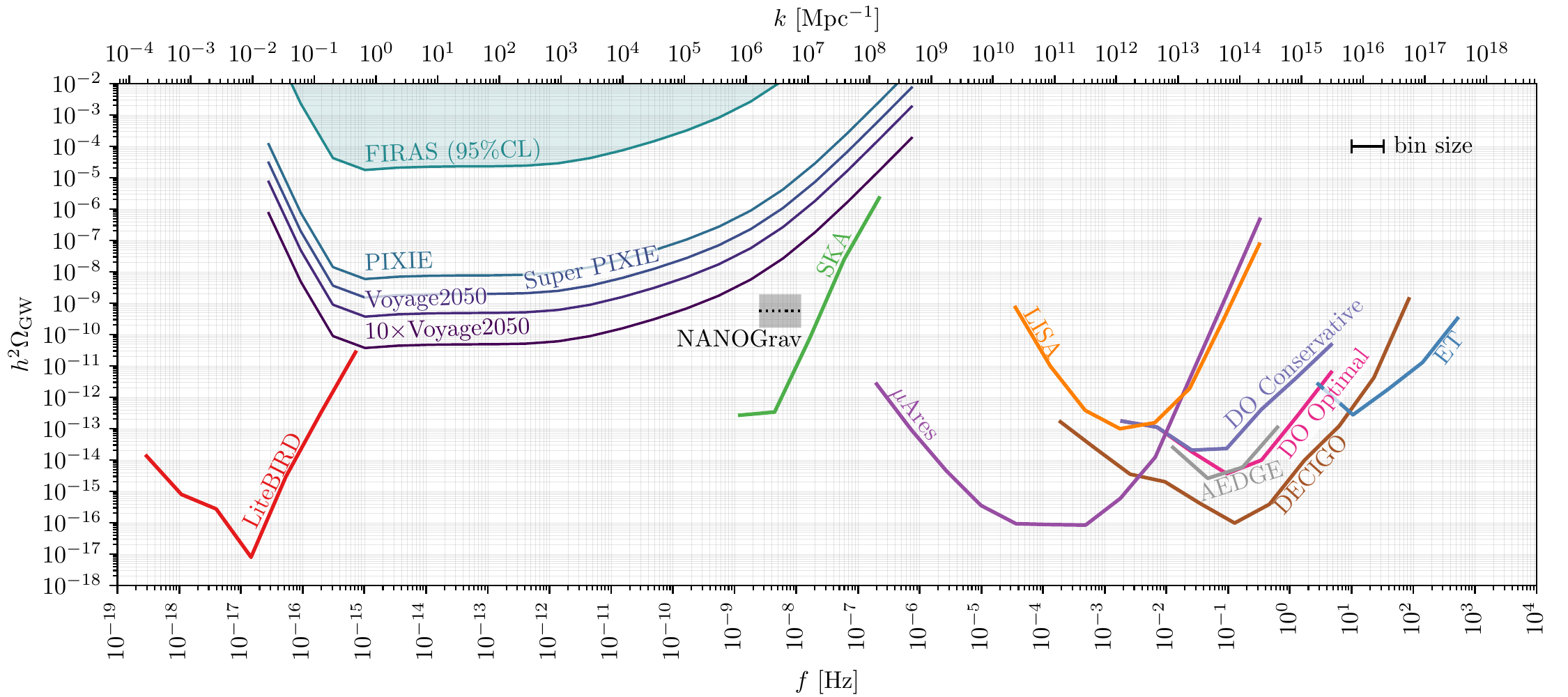}
\vspace{-5mm}
\caption{
Upper limits on the energy density of gravitational waves from measurements of $\mu$-distortions for various experimental configurations (\FIRAS, \PIXIE, \SuperPIXIE, \Voyage, $10\times$\Voyage).
For ease of comparison, we also report the upper limits for various other CMB, PTA and direct detection experiments \citep[taken from][]{Campeti:2020xwn}, 
and the NANOGrav 12.5 years 95\% confidence interval assuming a flat spectrum.
}
\label{fig:Omegak_SensitivityCurves}
\end{figure*}
In this section, we calculate the constraining power of spectroscopic CMB measurements in a minimally parametric fashion.
As in \cite{Campeti:2020xwn}, we parametrize the primordial tensor power spectrum using logarithmically spaced tophat functions centered around some $\ln k_i$ with $ \ln k_{i+1} - \ln k_i = 1.2 \, \forall i$.
This allow us an easy comparison with Fig.~8 of their paper:
\begin{subequations}
\begin{gather}
    \mathcal{P}_T(k)
    =
    \sum_i A_i W_i(k) \, ,
\label{eq:Binned_PT}
\\
    W_i(k)
    =
    \begin{cases}
        1 & \text{if } \ln k \in [\ln k_i - 0.6, \ln k_i + 0.6]
    \\
        0 & \text{otherwise}
    \end{cases}.
\end{gather}
\end{subequations}
Therefore, for each $i$, we insert Eq. \eqref{eq:Binned_PT} into Eq. \eqref{eq:mu_GW}, and calculate the maximum value of $A_i$ that is compatible with the chosen $\langle \mu_\text{GW} \rangle (\eta_0)$ upper limit.
With that information we then use Eq. \eqref{eq:OmegaGW_PT_relation} to calculate the corresponding $\Omega_\text{GW}$ constraint.

In Fig. \ref{fig:Omegak_SensitivityCurves} we show the sensitivity curves for \FIRAS, \PIXIE, \SuperPIXIE, \Voyage and $10\times$\Voyage, which all include estimated penalties from foregrounds.
For comparison, we also report the sensitivity curves from
\cite{Campeti:2020xwn}, which recently compiled the results of many planned experiments
\citep{
Hazumi:2019lys,
Smith:2019wny,
Sedda:2019uro,
Sesana:2019vho,
Kuroyanagi:2014qza,
Crowder:2005nr,
Bertoldi:2019tck,
Reitze:2019iox,
Hild:2010id,
Bull:2018lat%
}.
Moreover, we show the NANOGrav 12.5 year observation \citep{Arzoumanian:2020vkk}, interpreted as GW stochastic background according to their 5 frequency power-law model \citep[see ][for more discussion on whether the signal can be inflationary]{Kuroyanagi:2020:bluePT}. Since the extrapolation of a red spectra would be favourable for a SD detection, we conservatively assume a flat spectrum.

While the existing constraint derived from the \FIRAS data is a few order of magnitude higher than other probes,
next generation satellites will start to bridge nicely the frequency gap existing between CMB observation and direct GW detection.
It is interesting to notice that the upper bound from SDs will cover a very broad range of frequencies (more than 5 decades in $f$).
As such, any signal that is not sharply peaked in frequency will generate a comparatively higher $\mu$-distortion, tightening the constraints on specific parametric models, as we will see in the next section.

\vspace{-2mm}
\section{Constraints on specific models}
In this section, we consider concrete models that generate GWs over a wide range of scales. 
For each of the following models it is enough to insert their corresponding tensor power spectrum into Eq.~\eqref{eq:mu_GW} or \eqref{eq:mu_GW_time_dep} to obtain the predicted $\mu$-distortion, depending on whether the injection is primordial or happens after reheating.

Generally speaking, once accounting for the limits on $r$ from \Planck \citep{Ade:2018gkx, Aghanim:2018eyx}, we understand that appreciable SDs can only be created by models with substantially enhanced tensor power at small scales. To also avoid future constraints at small scales, models with localized features at $f=10^{-15}-10^{-9}\,\Hz$ are most promising. In the context of PTs, for example, this identifies low-scale dark or hidden sector transitions at energies $\simeq 10$ MeV - $10$ eV in the post-inflation era as a target.

\vspace{-3mm}
\subsection{Single-field slow-roll inflation}
As a benchmark we consider the tensor perturbations generated by single-field slow-roll inflation.
This model predicts a very low, almost scale invariant tensor spectrum, and as such we cannot expect SD constraints to be competitive with either CMB measurements or future direct detections at small scales.
We however include the model for completeness, and as a point of comparison. The tensor spectrum from this model is given by
\begin{equation}
    \mathcal{P}_T^\text{ sf}
    =
    A_T
    \left(k/k_0
    \right)^{n_T},
\label{eq:PT_SingleField}
\end{equation}
where the amplitude of tensor and scalar perturbations $A_T$ and $A_S$ are related by the tensor to scalar ratio by $r\equiv A_T/A_S$, and $n_T=-r/8$ \citep{Lyth:1998xn}.
Current constraints, mostly driven by \Planck low-$\ell$ temperature and BICEP2/Keck $B$-modes data, \citep{Ade:2018gkx, Aghanim:2018eyx} set the upper limit $r_{0.002} < 0.06 \ (95\%)$ at $k=0.002\,\text{Mpc}^{-1}$. Upon noticing $|n_T|\leq0.0075\approx 0$, one can approximate the result by integrating a flat spectrum yielding $\langle\mu\rangle(r)\approx\pot{1.68}{-13}\,r$ which gives the correct result to within $\leq 5\%$ for all values not ruled out by \Planck. This shows that for any allowed value of $r$ the SD signal will be out of reach for even the most sensitive SD mission concepts.

In principle this contribution is present as a component of tensor spectrum in the other models considered in the following sections. However, since the amount of SD it generates is anyway negligible, we will omit it in the following.
Note that the \Planck constraint on $r$ will also be considered for other models.
Strictly speaking, the aforementioned constraint only apply to a power-law tensor spectrum, a condition not necessarily met by the models we will consider in the following.
To provide some context to the SD constraints we will draw, we opt to employ an order-of-magnitude estimate of the \Planck constraint, simply requiring that any spectrum of tensor perturbations, $\mathcal{P}_T(k)$, must satisfy $\left.\mathcal{P}_T(k)/\mathcal{P}_S^{\rm\,sf}(k)\right\rvert_{k=0.002\text{ Mpc}^{-1}} < 0.06$.
In principle, a proper analysis of the \Planck and BICEP2/Keck data could be carried out to set constraints on the models that will be discussed here. This, however, goes beyond the scope of the paper.
{\it LiteBIRD} \citep{Hazumi:2019lys}, providing low multipole $BB$ information at much higher precision, will allow us to further improve the limits set by \Planck on the same range of scales in the near future.

\vspace{-3mm}
\subsection{Spectator SU(2) axions}
Many inflationary models require the dynamics of additional spectator fields active during the inflationary period, itself driven by a separate scalar field. Generally speaking, the dynamics of the spectator field generate tensor perturbations in addition to those produced by the vacuum fluctuations of the quasi de Sitter background.

In this section, following \cite{Campeti:2020xwn}, we consider an axion-SU(2) spectator field based on the ``chromo-natural'' inflation model \citep{Adshead:2012kp, Dimastrogiovanni:2016fuu}.
Here, the SU(2) gauge fields acquire an expectation value, the fluctuations around which include a tensor perturbation with a bilinear coupling to the graviton. The dynamics of the spectator SU(2) axion are such that gravitons of a particular helicity are amplified via a transient tachyonic instability, resulting in a (circularly polarized) contribution to the tensor power spectrum.

The spectrum for this model is given by \citep[see][]{Thorne:2017jft}
\begin{equation}
    \mathcal{P}_T^\text{ SU(2)}(k) 
    =
    r_* \mathcal{P}^\text{ sf}_\mathcal{R}(k) 
    \exp \left[
        -\frac{1}{2\sigma^2} \ln^2\left(\frac{k}{k_p}\right)
    \right] ,
\label{eq:PT_AxionSU2}
\end{equation}
which relates to the spectrum of scalar perturbations
\begin{equation}
\mathcal{P}^\text{ sf}_\mathcal{R}
    =
    A_S
    \left(
    k/k_0
    \right)^{n_S-1}.
\label{eq:PR_SingleField}
\end{equation}
In order to constrain this model, we use the best-fit \Planck parameters \citep{Planck2018params} for Eq.~\eqref{eq:PR_SingleField}. However, $r_*$, $k_p$ and $\sigma$ are related to the parameters of the gauge theory and are essentially free to vary here.
We take as a reasonable set of values, those given in \cite{Campeti:2020xwn} (their model AX3)
$(r_*, k_p, \sigma) = (50, 10^6 \text{ Mpc}^{-1}, 4.8$), which would yield $\mu=\pot{2.1}{-12}$. Entertaining the question as to which set of parameters would maximize the $\mu$ distortion signal while satisfying both observational and model constraints we find $(r_*, k_p, \sigma) =(265, 2.85\times10^4 \text{ Mpc}^{-1}, 4.02)$. However, even this best case scenario leaves no appreciable SD signal, yielding $\mu=\pot{2.1}{-11}$. This result can be understood by considering the parabolic shape of the spectrum in log$\mathcal{P}$-log$k$ space, which due to model constraints cannot peak to sharply \citep[e.g. see Eqs. (A8) and (A11) in][]{Thorne:2017jft}. This means that a spectrum which avoids the \Planck constraints cannot simultaneously peak too high in the SD regime.
In contrast, the models considered in the following subsections have spectra resembling broken power laws, and can be much more effective in satisfying current constraints while simultaneously generating significant SDs.

\vspace{-3mm}
\subsection{Ultra-light U(1) audible axions}
\label{sec:audit_axion}
The sensitivity of SD measurements to axion standard model extensions have already been discussed in the literature \citep{Mukherjee:2018:axionSD}. In this subsection however we consider a model proposed in \cite{Machado:2018nqk, Machado:2019xuc}, in which the axions specifically produce a strong GW signal. In this scenario (generic in the context of string compactifications), one has the presence of one or more U(1) axions with mass $m$ and decay constant $f_\phi$ that couple to dark sector photons. At early times during radiation domination, when the Hubble parameter $H$ is greater than $m$, the axion field is over-damped and is frozen. Once $H \lesssim m$, corresponding to the temperature $T \approx \sqrt{m M_{\rm pl}}$, the axion starts to oscillate around the minimum of its potential, sourcing gauge field production of a particular helicity that goes on to generate GWs. Since these GWs are only produced on sub-horizon scales after the axion starts oscillating, the results of Sect. \ref{sec:time_dep_injection} are essential in finding the $\mu$ signal accurately. 

The audible axion scenario features a qualitative difference with models that secondarily generate gravitation waves via gauge field production during inflation, as the generation occurs during radiation domination, when $H < m$. The oscillating axion at the minimum of the potential must remain a sub-dominant contribution to the energy budget, otherwise we'd have a phase of intermediate matter domination.
\newpage
It follows that the contribution of the oscillating axion and the subsequently produced dark photons must be sufficiently sub-leading to the energy density in the radiation fluid. Their relative contributions to the curvature perturbation in total comoving gauge will consequently be suppressed relative to the contribution from fluctuations in the radiation fluid already present before dark photon production (originating from the vacuum fluctuations sourced during inflation). Hence the contribution of scalar perturbations sourced by axion dynamics to the $\mu-$distortion signal will be sub-leading to those generated by primordial perturbations from inflation, and can safely be neglected. 

This model is of particular interest to us as it produces a narrower spectrum of GWs. Thus, to constrain its parameter space it is important to have probes that can cover all phenomenologically relevant frequencies. The GWs produced can be parametrized as a spectrum of the form 
\begin{subequations}
\begin{gather}
    \Omega_{\rm GW}^\text{ U(1)}(k) 
    =
    \frac
    {6.3 \Omega^\text{ U(1)}_\text{GW}(f_\text{AA}) \left(k/\tilde{k} \right)^{1.5}}
    {1+\left(k/\tilde{k} \right)^{1.5} \exp\left[12.9 \left(k/\tilde{k} -1 \right) \right]}\, ,
\label{eq:PT_AxionU1}
\end{gather}
with
\begin{gather}
    \tilde{k} = 1.3 \times 10^{15} \, \left[f_\text{AA}/\text{Hz}\right] \, \text{Mpc}^{-1} .
\end{gather}
\end{subequations}
Here $\Omega_\text{GW}(f_\text{AA})$ and $f_\text{AA}$ are a function of the free parameters of the model.
These parameters, as introduced in \cite{Machado:2018nqk,Machado:2019xuc}, are $f_\phi$, $m$, $\alpha$ and $\theta$, relating to the \textit{fit parameters} in Eq. \eqref{eq:PT_AxionU1} via
\begin{subequations}
\begin{gather}
f_{\rm AA}\approx 6\times 10^{-4}\,\Hz\,\left[\frac{\alpha\theta}{66}\right]^{2/3}\left[\frac{m}{10\text{meV}}\right]^{1/2},
\\[1mm]
\Omega^\text{ U(1)}_{\rm GW}(f_{\rm AA})\approx 1.67\times 10^{-4}g_{\rho,*}^{-1/3}
\left[\frac{f_\phi}{M_{\rm pl}}\right]^4
\left[\frac{\theta^2}{\alpha}\right]^{4/3}
\, ,
\end{gather}
\end{subequations}
which have both been redshifted to their present-day values.
The first two free parameters (i.e., $f_\phi$ and $m$) essentially dictate the height and frequency of the peak in the power spectrum respectively. The second two parameters are limited to $\alpha\sim 10-100$ and $\theta\sim\mathcal{O}(1)$, and do not significantly change the shape of the spectrum for the range of allowed values. These parameters are therefore degenerate with the first two. We choose fiducial values of $\alpha=60$ and $\theta=1$, but the main results given here hold more generally.

\begin{figure}
\centering 
\includegraphics[width=\columnwidth]{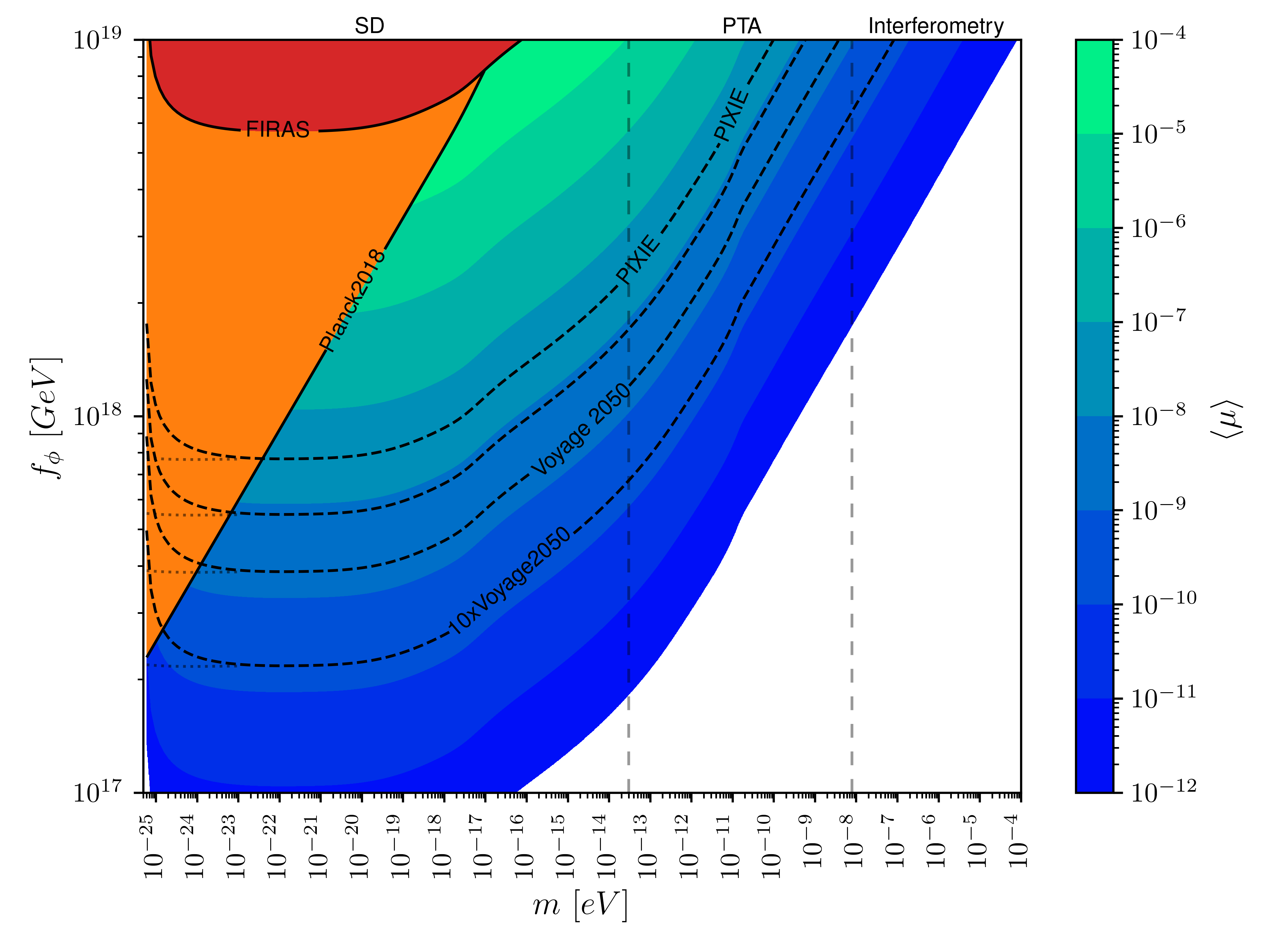}
\caption{A contour plot showing the expected SD signal arising from different combinations of $f_\phi$ and $m$ in the U(1) model. Without loss of generality, fiducial values of $\alpha=60$ and $\theta=1$ were chosen. Contours showing the visibility of several proposed spectrometers are shown. Vertical dotted lines indicate regions of the phase space where different probes are most sensitive (from left to right: SD, PTA and Interferometry). Dotted lines continuing the SD mission contours show the estimates ignoring late time injection.}
\label{fig:U1_contour}
\end{figure}

The direct dependence of $f_{\rm peak}$ on $m$ means that different types of experiment will probe different mass scales. This is shown Fig.~\ref{fig:U1_contour}, where vertical dotted lines distinguish where different detection methods are dominant. From here it can be seen that SD are sensitive to the ultralight limit of the $U(1)$ audible axion model, a result which again holds for any valid combination of $\alpha$ and $\theta$.

Note that \Planck extends the limits from \FIRAS at low masses to smaller values of $f_\phi$. Future SDs measurements could significantly improve the limits from \Planck to higher masses, covering a wider range of the parameter space of phenomenological interest. 

We note in particular how SDs can constrain masses in a range not accessible to other measurements ($10^{-22} - 10^{-13}$ eV). Such ultra-light axions may be ubiquitous in particular string compactifications \citep{Arvanitaki:2009fg}, and moreover, could be a viable dark matter candidate were they to form a condensate at late times \citep{Hui:2016ltb, Marsh:2015xka}, further illustrating the utility of SDs for particle phenomenology. 
\newpage
\vspace{-3mm}
\subsection{Phase transitions beyond the Standard Model}
\label{sec:phase_transition}

The post-inflationary epoch may have seen a variety of first order phase transitions  (PTs) in theories that go beyond the standard model of particle physics. First order PTs are characterized by the fact that latent energy is released, and phases of true vacuum nucleate within false vacuum domains, resulting in bubble collisions (BC) that generate a stochastic GW background. Moreover, magneto-hydrodynamic (MHD) turbulence and sound waves (SW) in the bulk plasma during and after the phase transition also source sub-horizon GWs at commensurate frequencies. If these processes take place during the $\mu$-era or shortly before, they can potentially result in measurable SDs. Here we once again use the results of Sect.~\ref{sec:time_dep_injection} to calculate the associated SDs.

Referring to the review of \cite{Caprini:2018mtu}, we see that the spectra resulting from the three different mechanisms for GW production from PTs are given by
\begin{subequations}
\begin{align}
h^2\Omega^{\rm BC}_{\rm GW}(f) = \,\,&
1.67\times 10^{-5}
\left(\frac{H_*}{\beta}\right)^2
\left(\frac{\kappa_{\rm BC}\alpha}{1+\alpha}\right)^2
\left(\frac{100}{g_*(T_*)}\right)^\frac{1}{3}\\
& \times \left(\frac{0.11\varv_w^3}{0.42+\varv_w^2}\right)
\frac{3.8(f/f_{\rm BC})^{2.8}}{1+2.8(f/f_{\rm BC})^{3.8}}\,,\nonumber\\
h^2\Omega^{\rm SW}_{\rm GW}(f) = \,\,&
2.65\times 10^{-6}
\left(\frac{H_*}{\beta}\right)
\left(\frac{\kappa_v\alpha}{1+\alpha}\right)
\left(\frac{100}{g_*(T_*)}\right)^{\frac{1}{3}}\\
& \times \varv_w
\left(\frac{f}{f_{\rm SW}}\right)^3
\left(\frac{7}{4+3(f/f_{\rm SW})^2}\right)^{\frac{7}{2}}\,, \nonumber\\
h^2\Omega^{\rm MHD}_{\rm GW}(f) = \,\,& 3.35\times10^{-4}
\left(\frac{H_*}{\beta}\right)
\left(\frac{\kappa_{\rm MHD}\alpha}{1+\alpha}\right)^{\frac{3}{2}}
\left(\frac{100}{g_*(T_*)}\right)^{\frac{1}{3}}\\
& \times \varv_w
\frac{(f/f_{\rm MHD})^3}{\left[1+(f/f_{\rm MHD})\right]^\frac{11}{3}\left(1+8\pi f/h_*\right)}\,,\nonumber
\end{align}
with peak frequencies
\begin{align}
\chi_0=\,\,&\left[\frac{\beta}{H_*}\right]
\left[\frac{T_*}{100{\rm GeV}}\right]
\left[\frac{g_*(T_*)}{100}\right]^\frac{1}{6}\,,\\[1mm]
f_{\rm BC}=\,\,&1.65\times 10^{-5}\,{\rm Hz}
\,\left(\frac{0.62}{1.8-0.1\varv_w+\varv_w^2}\right)\,\chi_0\,, \\
f_{\rm SW}=\,\,&1.9\times 10^{-5}\,{\rm Hz}
\,\varv_w^{-1}\,\chi_0\,,\\
f_{\rm MHD}=\,\,&2.7\times 10^{-5}\,{\rm Hz}
\,\varv_w^{-1}\,\chi_0\,.
\end{align}
\end{subequations}
\begin{figure*}
\centering
\includegraphics[width=\linewidth]{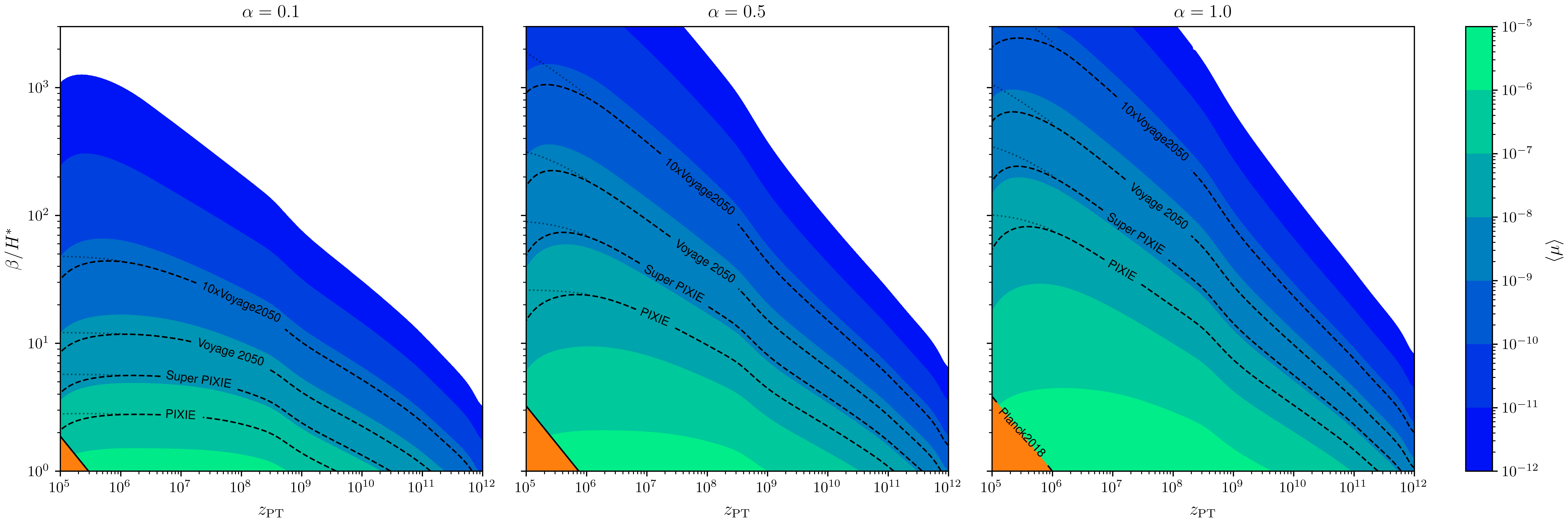}
\caption{A series of contour plots showing the expected SD signal arising from low scale first order phase transitions. Dotted lines (visible on the left) give the sensitivity with standard window function, $W_\mu(k)$, showing that late power injection leads to a decrease of less than an order of magnitude for $10^5<z_{\rm PT}<10^6$. The limits from \Planck are also shown for comparison. The temperature at the time of the PT can be found with $T_{\rm PT}/\text{MeV}\approx\pot{2}{-10}z_{\rm PT}$.}
\label{fig:PT_contours}
\end{figure*}
Here, the three principal model parameters are $\alpha$, $\beta$ and $\varv_w$, which fix the amount of latent heat released by the transition as a fraction of the total energy density, inverse time duration of the PT, and velocity of bubble walls respectively. Denoted with $*$ are quantities at the time of the PT, making another key parameter $z_{\rm PT}$.
The first two parameters follow $0\leq\alpha\leq 1$ and $\beta/H_*>1$.

The velocity of sound waves has been set to unity, since bubble walls usually propagate close to the speed of light. Parameters labelled $\kappa_i\in[0,1]$ give the weighted contribution from each mechanism. For this work we have used $\kappa_{\rm BC}=1$, $\kappa_{\rm MHD}=\kappa_v$ and
\begin{equation}
\kappa_v \approx \frac{\alpha}{0.73+0.083\sqrt{\alpha} + \alpha},
\end{equation}
the last of which is valid for $\varv_w\simeq 1$.
The expected SD limits on PTs given these considerations are shown in Fig.~\ref{fig:PT_contours}. Even for low-energy PTs $(\alpha=0.1)$ a \PIXIE-like mission would explore some of the parameter space not already excluded by \Planck; however, it would only see rather long PT. In the more energetic cases $(\alpha\geq0.5)$, SD missions could realistically detect PT lasting small fractions of the age of the Universe, and occurring relatively late in cosmic history. 

Evidently, SDs provide a unique and complimentary window into low scale phase transitions (corresponding to energy scales in the range 10 Mev - 10 eV) that are not possible to probe with any other observation. An important caveat to our discussion of this scenario is the potential for the generation of sub-horizon scalar perturbations during and after phase transitions. Sub-dominant contributions arising from the scalar field dynamics have been calculated in \cite{Cutting:2018tjt}, however retaining the scalar contributions from sound waves and MHD turbulence generated after the transition will require further study, and remains an important open question for the present analysis. Our limits can therefore be considered conservative.

\subsection{GUT cosmic string networks}
Another tell tale sign of physics beyond the Standard Model is the existence of topological defects. Excluding textures, the standard model does not allow for any defects. However, larger gauge symmetries (ubiquitous in models that go beyond the Standard Model) could admit symmetry breaking patterns that generate topological defects in the early Universe \citep[see][]{Kibble:1980,Kibble:1982} which could have persisted into cosmologically observable epochs. Although the simplest models of monopoles and domain walls are tightly constrained \citep[see Sects.~13.5.3 and 14.3.3 in][]{Vilenkin:1994}, cosmic strings networks remain a theoretical possibility and can impart potentially observable GW signals \citep[see Sect.~10.4 of][]{Vilenkin:1994}.

As an example, we consider a model proposed by \cite{Buchmuller:2019:CSGW} which attempts leptogenesis within an SO(10) grand unified theory via a U(1)$_{B-L}$ phase transition, where a local U(1) baryon minus lepton number symmetry is spontaneously broken.
\newpage
The result of the $B-L$ transition will be a meta-stable CS network generated at the time of the transition, which over the course of the collapse generates a mostly flat spectrum of GWs due to the decay of string loops. An approximate form of their spectrum is given in terms of the model parameters $\kappa$ and $G\mu$ as\footnote{Not to be confused with the SD amplitude $\mu$. The combination $G\mu$ will always be in reference to the energy scale of the CS physics.}
\begin{subequations}
\begin{gather}
    h^2\Omega^\text{ CS}_{\text GW} = h^2\Omega_{\rm GW}^{\rm plateau}\text{min}\left[\left(f/f_*\right)^{3/2},1\right]\, ,
    \\[1mm]
    f_*=\pot{3}{14} \,\Hz\, {\rm e}^{-\pi\kappa/4}\left[\frac{G\mu}{10^{-7}}\right]^{-1/2}\, ,
    \\
    h^2\Omega_{\rm GW}^{\rm plateau}=8.04\,\Omega_{\rm r}h^2\,\left[\frac{G\mu}{\Gamma}\right]^{1/2}\,.
\end{gather}
\end{subequations}
\cite{Buchmuller:2019:CSGW} give a value of $\Gamma\approx 50$ for this particular model, and we use a value of $\Omega_{\rm r}h^2=2.5\times 10^{-5}$.

In reality string network collapse would be a function of time, but to match the formalism outlined in Sect. \ref{sec:time_dep_injection} we conservatively assume the entire spectrum emerges at the final moment of collapse given by \cite{Buchmuller:2019:CSGW}
\begin{equation}
    z_{\rm collapse}=\left(\frac{70}{H_0}\right)^{1/2}
    \left(\Gamma\,\frac{(G\mu)^2}{2\pi G}e^{-\pi\kappa}\right)^{1/4}.
\end{equation}
\begin{figure}
\centering
\includegraphics[width=\linewidth]{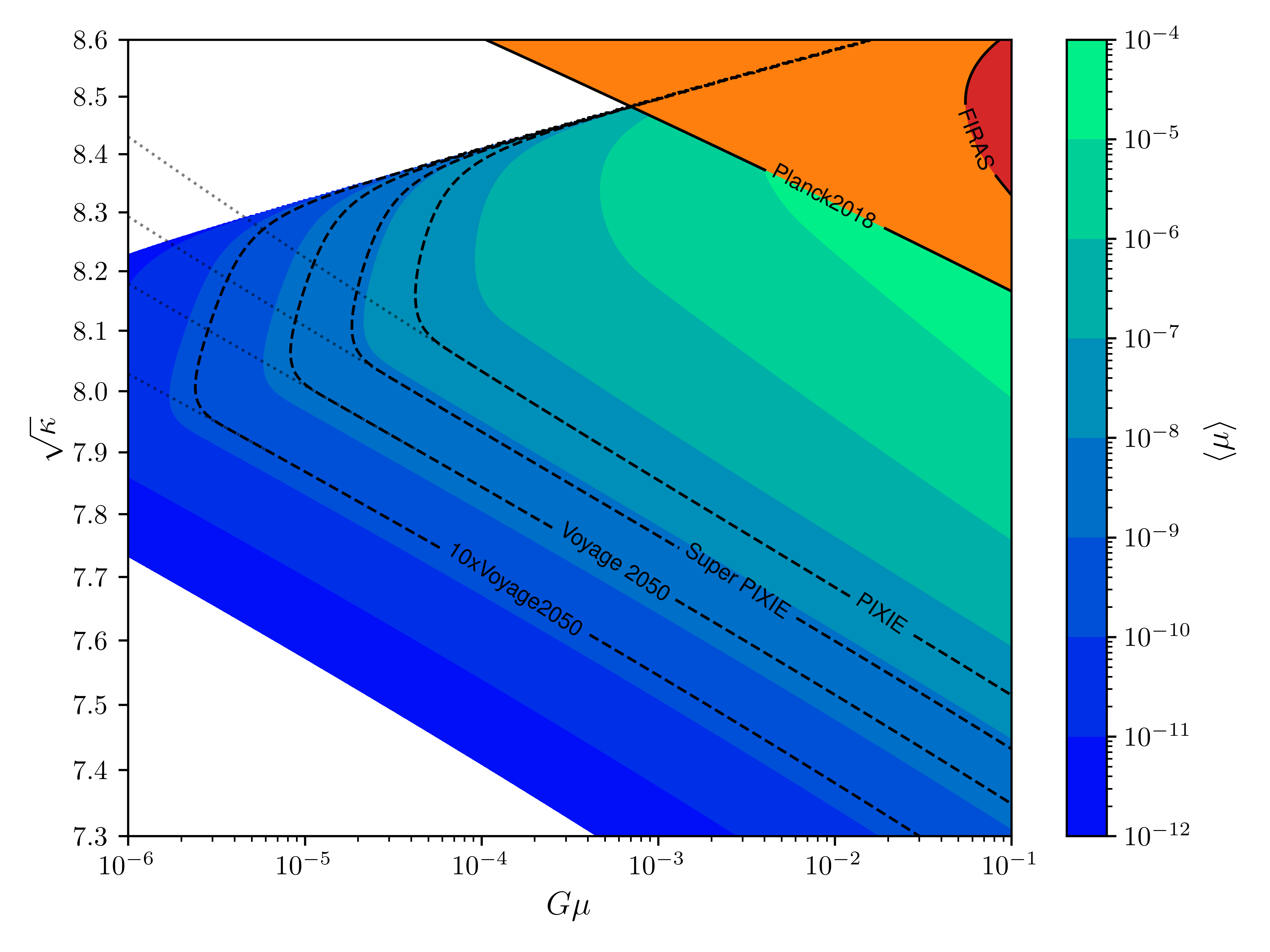}
\vspace{-2mm}
\caption{A contour plot showing the expected $\mu$ signal from a CS network arising from a U(1)$_{B-L}$ phase transition at the GUT scale. The limit placed by \FIRAS is shown in red, and similarly for \Planck in orange. Dashed contours show the sensitivity of various proposed SD missions. Faint dotted lines show the contours without using the $z_{\rm max}$ limited window function.}
\label{fig:CS_network}
\end{figure}
The spectrum grows $\propto f^{3/2}$ up to $f_*$, and is flat for higher frequencies. Furthermore, $f_*$ only depends weakly on $G\mu$ but varies significantly with $\kappa$. This means that once $\kappa$ is large enough that the spectrum is flat across the entire window of visibility for a given experiment, the probe will only be sensitive to $G\mu$. With SD missions probing lower frequencies than astrophysical probes they will be complementary in limiting the lower bounds of the $\kappa$ parameter. The potential of SD missions for constraining this model is shown in Fig.~\ref{fig:CS_network}. 

Given that the GW spectra produced by CS network collapse has a plateau at smaller scales, for any given sensitivity depicted in Fig.~\ref{fig:Omegak_SensitivityCurves}, we see that any one of the probes depicted will be equally good at detecting the GW background produced. It is also worth noting that the type of spectrum considered here will hold more generally for a wide range of CS models \citep[see][]{Figueroa:2020:CSGW}.
\newpage
Although the generation of scalar perturbations of CS networks in the scaling regime is well understood, the situation is much less clear for the scalar perturbations generated from the decay of meta-stable networks. Moreover, it is unclear whether the dominant decay channel will be via gravitational wave production or scalar perturbations, and the answer will certainly be very model dependent. This remains another open question as far as this study is concerned, and the absence of any such detailed calculations is itself perhaps due to the fact that it may have been unclear in the past what observational consequences, if any, sub-horizon generation of scalar perturbations generated by CS network collapse would have. We hope the results of this paper will provide the necessary motivation for such calculations.
\newpage
\section{Discussion and conclusions}
Highly energetic events in the early Universe, either during inflation or subsequently during radiation domination, can inject power into the GW spectrum. This can include GWs from sources within the standard $\Lambda$CDM cosmology, or from models that invoke physics beyond it. Detecting the GW spectrum is therefore key to further scrutinizing our current paradigm, as well as pushing our knowledge of the early Universe to new and exciting areas. Future experiments will probe these stochastic backgrounds, each sensitive to a  range of frequencies/wavelengths dictated by the nature of the experiment. 
As we have highlighted here, a wide range of GW frequencies (\mbox{$f=10^{-15}$--$10^{-9}\,\Hz$})
can only be probed by SD observations. This large span of wavelengths compensates for the relatively low efficiency of generating SDs from GWs, thus making them a potentially powerful probe of physics beyond the Standard Models of both particle physics and cosmology.

This work aims to introduce SDs as a complimentary probe through which one can detect and constrain stochastic GW backgrounds. The fundamental element to link these two messengers is the $k$-space window function, which maps a given GW spectrum into a SD signal imprinted before last scattering [see Eqs.~\ref{eq:mu_GW} and \ref{eq:mu_GW_time_dep}]. In order to study the injection of power on sub-horizon scales, the window function for primordial tensor perturbations has been generalised [see Eq.~\ref{eq:mu_GW_time_dep}], leading to minor changes in some models (Fig. \ref{fig:U1_contour}) but large changes in others (Fig. \ref{fig:CS_network}). This is essentially related to the fact that GWs have less cosmic history to dissipate their energy to the photon-baryon plasma. A simple python tool is provided at \SDGWtool\footnote{\url{https://github.com/CMBSPEC/GW2SD.git}} and allows one to easily estimate SD limits on various models, given the tensor power spectrum, $\mathcal{P}_T(k, z)$, that comes into existence at a single redshift $z$. This is certainly a good approximation for 1'st order phase transitions, and holds to a good approximation for scenarios that dynamically generate GWs over a short duration.
Refinements to account for the exact time-dependence of the process are left to future work.

To illustrate the utility of SDs for GW cosmology, a series of phenomenological models were discussed, and their resulting SD signals studied:
As expected, the tensor perturbations generated by single-field slow-roll inflation are too weak to be measured with SDs \citep{Ota2014, Chluba2015}.
Spectator axion-SU(2) fields too, even in more favourable cases that we considered, will realistically be out of reach in the foreseeable future. 
The Audible axion model (Sect.~\ref{sec:audit_axion}) on the other hand, can have a large region of its parameter space constrained by SDs, particularly for a wide range of masses in the ultra-light regime (Fig.~\ref{fig:U1_contour}). Similarly, the GWs from low scale (10 eV - 10 MeV) dark sector phase transitions in the early Universe will be visible with future SD missions if the relative energy content of the participating field is sufficiently large, and the duration sufficiently long (see Sect.~\ref{sec:phase_transition} and Fig.~\ref{fig:PT_contours}). The typically flat GW spectra produced by CS networks can be seen by many instruments, but SDs will be complementary to other probes in being sensitive to string collapse especially in the $\mu$-era. It is noteworthy that some of the aforementioned models were already constrained with \FIRAS long before first limits from \Planck existed. Future CMB spectrometers like \SuperPIXIE \citep{Kogut2019BAAS} could establish a new frontier in this respect. 
\newpage
As we have discussed, both in general and for specific models, any SD constraints should include both the scalar and tensor perturbations arising from energetic events \citep[e.g. see][for SDs from scalar perturbations of CS and PTs, respectively]{Tashiro2012b, Amin2014}.
This potential for combining sources is another advantage SD experiments have over GW-based experiments, since the latter are only sensitive to the direct tensor perturbations. 
This advantage is not easily utilised for the models discussed in this paper, however, since the necessary scalar spectra are generally absent from the literature.
Where the inclusion \textit{is} possible, it is again important to highlight that SDs from tensor perturbations cover a wider range of physical scales than SDs from scalar sources, thus extending the reach of SDs to earlier epochs. In addition, some scenarios do not produce any significant scalar perturbations [e.g., the axion-SU(2) model], making it crucial to account for SDs caused by tensor perturbations. Overall, SDs uniquely probe the presence of small-scale perturbations in regimes that are not directly accessible, thus highlighting the important role that future CMB spectrometers could play in GW cosmology, and, by extension, beyond the Standard Model phenomenology. 

\section*{Data Availability}
Window functions (e.g. Fig. \ref{fig:z_max_window}) are available at \url{https://github.com/CMBSPEC/GW2SD.git}.

{\small
\section*{Acknowledgments}
We would like to thank Ana Achucarro and Jose Juan Blanco-Pillado for discussions about scalar perturbations generated during cosmic string collapse.
We also thank Paolo Campeti and Eiichiro Komatsu for providing the data used in Fig. \ref{fig:Omegak_SensitivityCurves}, together with helpful discussions regarding the mapping between $\mathcal{P}_T$ and $\Omega_{\rm GW}$.
This work was supported by the ERC Consolidator Grant {\it CMBSPEC} (No.~725456) as part of the European Union's Horizon 2020 research and innovation program.
TK was further supported by STFC grant ST/T506291/1.
JC was also supported by the Royal Society as a Royal Society URF at the University of Manchester, UK.
}

{\small
\bibliographystyle{mn2e}
\bibliography{bibliografia,Lit}
}

\end{document}